\documentclass[letterpaper,12pt]{article}   
\usepackage{osajnl2} 
\usepackage[draft]{hyperref} 

\begin{document}


\title{Full three dimensional inverse design of dielectric slab based scattering optical elements devices}


\author{Andreas H\aa kansson,$^{1,*}$ Stefano Boscolo,$^2$ and Michele Midrio$2$}
\address{$^1$NIMS, International Center for Young Scientists, 1-1 Namiki, Tsukuba, Ibaraki 305-0044 JAPAN}
\address{$^2$INFM, Dipartimento di Ingegneria Elettrica, Gestionale e Meccanica, Universit\`a degli Studi di Udine, Udine 33100, ITALY}
\address{$^*$Corresponding author: ahakansson@dasphotonics.com}

\begin{abstract}
Inverse design through optimization of dielectric photonic devices is a very powerful tool. However so far the direct solver used in the design process is almost always restricted to solve Maxwell's equation in two dimensions (2D). Here we will show that by using a specific three dimensional (3D) electromagnetic field-solver we can implement a full 3D inverse design tool for Silicon On Insulator (SOI) slab based photonic structures which can be run on a normal desktop computer.
\end{abstract}



\maketitle 

Optimization and automatic design of photonics components is a very desirable goal in many fields in photonics. As more and more sophisticated fabrication methods are evolving for dielectric material processing, the design space becomes more and more open. By exploiting this perfection it has been shown that classical solutions are not guaranteed to be the optimal solutions. For example, unconventional dielectric waveguides bends with low losses have been designed and measured\cite{ref:PIBorelOE04} and perfect light emitting cavities have been proposed\cite{ref:AHakanssonPRL06}. In addition, in the microwave regime it has been demonstrated how to fully control the light propagation through multiple scattering of rods in air\cite{ref:PSeligerJAP06}. Then again, by using the same method, a classical solution can also be proven to be the actual optimal solution. This was shown by Gondarenko et. al when designing a photonic cavity, by starting from random material distribution their design converged to a conventional Bragg mirror cavity\cite{ref:LipsonPRL06}.

The inverse design (ID) tool in optics is assembled by an electromagnetic field solver and a optimization algorithm. As for 2D photonic device designs, the multiple scattering technique (MST) \cite{ref:AIshimaruEMWPRS} in conjugation with a stochastic search algorithm, the genetic algorithm (GA) \cite{ref:DEGoldbergBookGA}, has shown great potential \cite{ref:LSanchisAPL04, ref:AHakanssonAPL05, ref:AHakanssonOE05, ref:AHakanssonPRL06}. The 2D MST-GA tool is using a simple analytic direct solver for simulating light propagation in clusters of similar sized dielectric scatterers with a diameter of the size of the wavelength or less. The optimal distribution of the scatterers with respect to a fixed functionality is found by optimizing a fitness function which directly represents the qualitative quantity of a devices performance.

An ID process is very computational costly since many configurations have to be simulated before the final result is obtained. More general and popular solvers like finite different time domain (FDTD) is therefore not a desirable choice. We will here show how a very specific direct solver, i.e. the 3D MST, can be used in order to implement a fast and accurate 3D-ID tool. This semianalytic EM solver, developed by some of us, can simulate the EM field distribution for dielectric slab structure much faster than conventional FDTD tools. For details please refer to the paper by Boscolo and Midrio \cite{ref:SBoscoloJLT04}.

In this paper we treat the propagation in a scattering optical elements (SOE) structure realized in a 250nm thick Si (n=3.4) membrane suspended in air. When the light propagates, confined in the guiding high refractive index slab, the scattering is controlled by carefully optimizing the position of the scattering elements, i.e. the etched holes in the slab. Figure 1 shows a cartoon of the 3D-MST computational domain that includes a multilayer structure constituted of a high refractive index guiding layer, two air layers and two perfectly matched layers as boundary conditions.

Due to the fact that these scattering objects are geometrically simple it is possible to rigorously solve Maxwell's equations in analytically form. To express the field inside and around each cylindrical object, the best choice is to write the field as a sum of the natural modes, the cylindrical harmonics. With this choice there is no need to make a space or time discretion, and few terms will be sufficient to approximate the field. The 2D-MST has widely been used in ID since it is one of the most powerful, fastest, and most accurate simulation tool for electromagnetic 2D calculations. About the choice of optimization tool, we are here using a GA, however the aim of this work is to show that the aforementioned simulation technique is well suited for specific ID problems. Hence, the discussion regarding the choice of optimization process will be excluded in this letter.

To test and analyze the 3D ID-tool we have chosen common device problem: A lens based on the SOE technology introduced in Ref. \cite{ref:AHakanssonAPL05}. We have chosen to limit the design space to air holes with a diameter of 200nm placed in a fixed square grid with a lattice parameter of 500nm. The total grid size is 6.0$\mu$m $\times$ 2.5$\mu$m equaling 65 lattice sites (LS). To control the absence or presence of a hole at a specific lattice site a binary design variable was addressed to each LS. However, since only symmetric solutions were considered it resulted in 35 binary variables. The complete design space equals a total of $2^{35}=3.4\times10^{10}$ different solutions. 

Which lattice sites that should be occupied by a air hole is decide by using the GA and maximizing the quality (fitness) value. The lens problem is defined by finding the optimal hole distribution that will harvest the highest intensity in a set point in space. The fitness of each device is simply evaluated by calculating the intensity in this focus for an incident TM-polarized Gaussian beam with a diameter of 6$\mu$m and of the wavelength $\lambda$=1550nm. Since the fitness value is set by a single field point, the computational speed is considerable fast. 
A full 3D fitness evaluation for a device including a distribution of about 35 holes (50\% of the lattice sites occupied) takes approximately 6s on a Pentium IV processor while using no significant size of memory. It has to be mentioned that the computational time is proportional to N$^2$, where N is the number of scatterers. On the other hand, no symmetry is implemented at present for the calculation. Using a symmetry condition the number of scatterers included in the device could be doubled without increasing the calculation time. 
Here, a very interesting question arise: Whether a full 3D solver is necessary for these geometries or a 2D solver would be able to solve the problem with sufficient accuracy and doing so much faster. The answer will of course depend on the design problem you want to solve. 
The lens problem treated here is a transparent device without light localization which should favor a 2D solver. Figure~\ref{fig:map3D} shows both 3D (upper) and 2D (lower) calculations of the best performing SOE lens designed using the full 3D-ID tool. The design was found after approximately 4000 evaluations or 6.5h hours calculation time. The fitness was calculated in the center of the Si membrane in the focal spot set on the symmetry axis at 8$\mu$m from the structure. The absolute intensity in the focus is $4.66 \times 10^{47}$ equaling an amplification of 7.0dBs. 

For comparison the same set-up was used in the 2D-ID program. In order to compensate for the finite thickness of the Si slab, the effective refractive index of the fundamental mode (n=2.82) was used in the 2D calculation. The simulation time for each fitness evaluation was approximately 1s per device. Since the optimization problem is identical, the number of evaluations needed before converging to the optimal design is the same as before (approx. 4000) equaling 1.2h of CPU time. This is about 5 times faster than the full 3D ID process. The final device can be seen in figure~\ref{fig:map2D}. The absolute intensity in the focus equals $4.27 \times 10^{47}$. Once again the 2D calculation has been included for comparison.

Hence, by using the full 3D calculation method the fitness of the device is increased by 9.1\% from 6.6dBs to 7.0dBs, if compared with the 2D-ID. On the one hand, this confirms an improvement of the performance. On the other hand, it also confirms that the 2D-ID tool is applicable. 
When dealing with bigger problems the time difference between the two solvers will increase due to non-linear calculation time behavior. Then it might be a good idea to consider using a 2D-ID design. 
However, other problems might include higher or lower differences. One very exciting problem is photonic cavity design. In SOI based photonic crystal cavities it is possible to confine the light in very small volumes with very high Q-factors by systematically tuning of the cavity\cite{ref:SNodaNAT03}. Using inverse design this tuning can be controlled by a sophisticated optimization process, addressing a much greater design space as well as controlling simultaneously suppression and enhancement of the spontaneous emission as was proposed in Ref. \cite{ref:AHakanssonPRL06}. 

In summary, a specific 3D-ID tool for dielectric layered structures has been presented. As a test problem, a photonic lens design has been addressed and analyzed by comparing the result from 2D design with full 3D design. 
This tool looks like a very promising candidate for photonic light emitting cavity design.

This study was performed through Special Coordination Funds for Promoting Science and Technology from the
Ministry of Education, Culture, Sports, Science and Technology (MEXT),

\clearpage

\section*{List of Figure Captions}

Fig. 1. Schematic cartoon of a SOE device from the SOI technoloy.

\noindent Fig. 2. The intensity response from a photonic lens designed using the complete 3D-ID tool. For comparison the scattered field was calculated using 3D MST (upper) and 2D MST (lower). The light propagates from left to right in the figure. The intensity is normalized to the maximum value in the focus.

\noindent Fig. 3. The intensity response from a photonic lens designed using the 2D-ID tool. For comparison the scattered field was calculated using 3D MST (upper) and 2D MST (lower). The light propagates from left to right in the figure. The intensity is normalized to the maximum value in the focus.


\clearpage


  \begin{figure}[htbp]
  \centering
  \includegraphics[width=8.3cm]{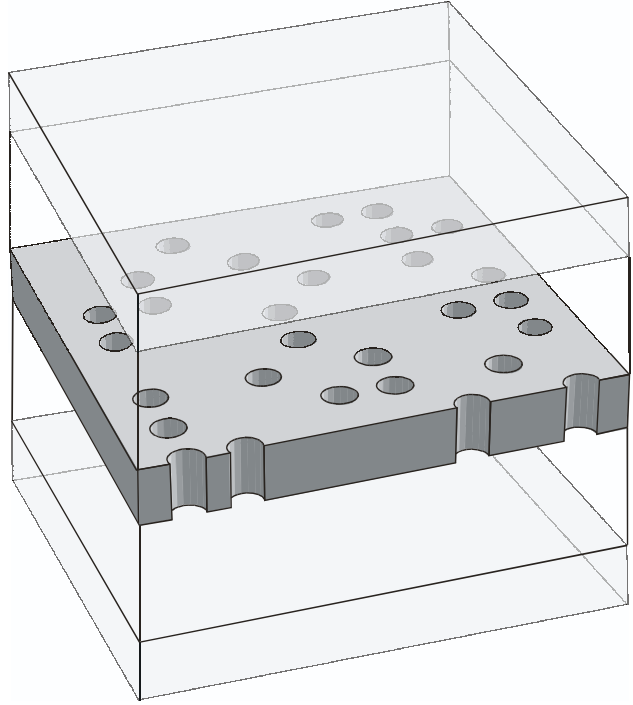}
  \caption{Schematic cartoon of a SOE device from the SOI technoloy.}
  \label{fig:illust}
  \end{figure}

\clearpage


  \begin{figure}[htbp]
  \centering
  \includegraphics[width=8.3cm]{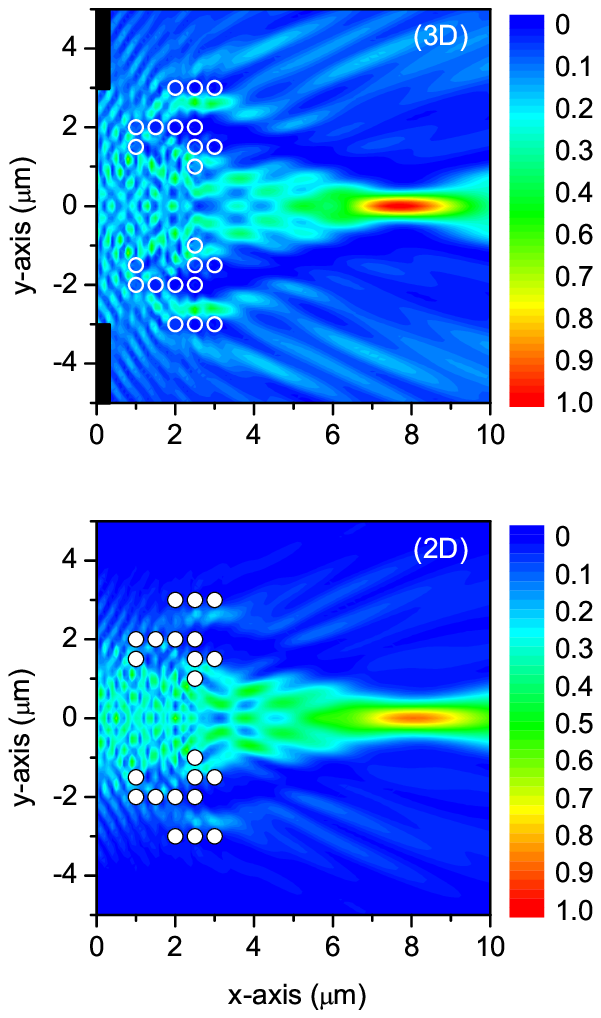}
  \caption{The intensity response from a photonic lens designed using the complete 3D-ID tool. For comparison the scattered field was calculated using 3D MST (upper) and 2D MST (lower). The light propagates from left to right in the figure. The intensity is normalized to the maximum value in the focus.}
  \label{fig:map3D}
  \end{figure}

\clearpage


  \begin{figure}[htbp]
  \centering
  \includegraphics[width=8.3cm]{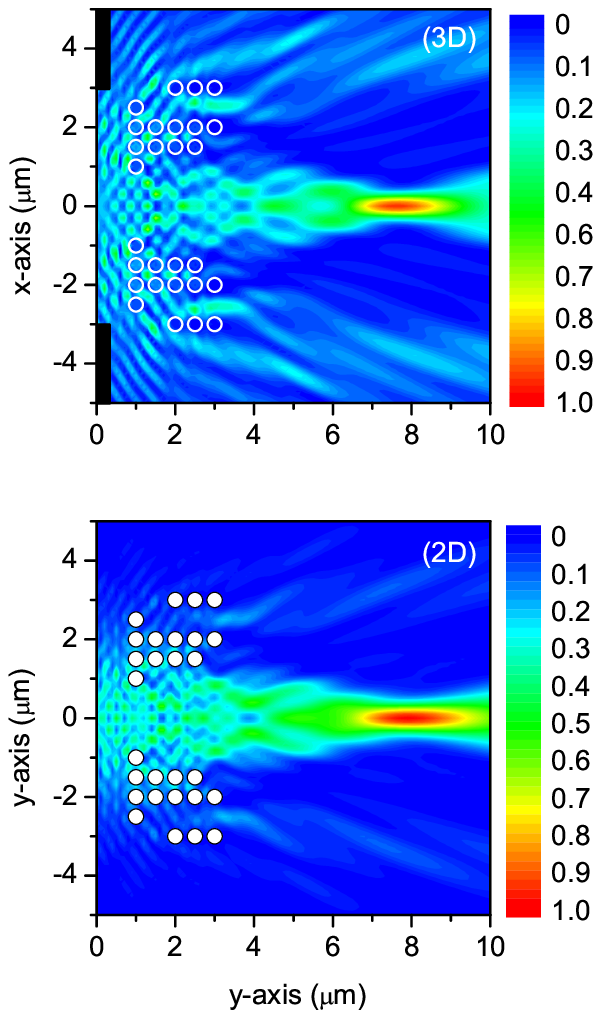}
  \caption{The intensity response from a photonic lens designed using the 2D-ID tool. For comparison the scattered field was calculated using 3D MST (upper) and 2D MST (lower). The light propagates from left to right in the figure. The intensity is normalized to the maximum value in the focus.}
  \label{fig:map2D}
  \end{figure}

\end{document}